\documentclass[pre,twocolumn,english,aps,a4paper,floatfix]{revtex4-2}
\usepackage[T1]{fontenc}
\usepackage{tikz}
\usepackage{color}
\usepackage{balance}
\usepackage{amsmath}
\usepackage{amssymb}
\usepackage[latin1]{inputenc}
\usepackage{graphicx}
\usepackage{bm}
\usepackage{epsfig}
\usepackage{xcolor}
\definecolor{mynicegreen}{RGB}{102,182,102}

\begin{document}

\title{Failure of standard density functional theory
to describe the phase behavior of a fluid of hard right isosceles triangles}

\author{Yuri Mart\'{\i}nez-Rat\'on}
\email{yuri@math.uc3m.es}
\affiliation{
Grupo Interdisciplinar de Sistemas Complejos (GISC), Departamento
de Matem\'aticas, Escuela Polit\'ecnica Superior, Universidad Carlos III de Madrid,
Avenida de la Universidad 30, E-28911, Legan\'es, Madrid, Spain}

\author{Enrique Velasco}
\email{enrique.velasco@uam.es}
\affiliation{Departamento de F\'{\i}sica Te\'orica de la Materia Condensada,
Instituto de F\'{\i}sica de la Materia Condensada (IFIMAC) and Instituto de Ciencia de
Materiales Nicol\'as Cabrera,
Universidad Aut\'onoma de Madrid,
E-28049, Madrid, Spain}

\date{\today}

\begin{abstract}
	A fluid of hard right isosceles triangles was studied using an extension of
	Scaled-Particle Density-Functional Theory which includes the exact third virial coefficient. We 
	show that the only orientationally ordered stable liquid-crystal phase predicted
	by the theory is the uniaxial nematic phase, in agreement with the second-order virial theory.
	By contrast, Monte Carlo simulations predict exotic liquid-crystal 
	phases exhibiting tetratic and octatic correlations, with orientational distribution 
	functions having four and eight equivalent peaks, respectively.
	This demonstrates the failure of the standard Density-Functional Theory based on two and three-body 
	correlations to describe high-symmetry orientational phases in two-dimensional 
	hard right-triangle fluids, and points to the necessity to 
	reformulate the theory to take into account high-order body correlations
	and ultimately particle self-assembling and clustering effects.
	This avenue may represent
	a great challenge for future research, and we discuss some fundamental 
	ideas to construct a modified version of Density-Functional Theory 
	to account for these clustering effects.
\end{abstract}

\maketitle

\section{Introduction}
\label{introduction}

The experimental realization of two-dimensional fluids of hard polygonal particles to study their phase behaviors
is an active line of research. For example, lithographic techniques applied to prepare nonoverlapping 
particles of a specific polygonal shape, and their adsorption or confinement, produce
single monolayers of Brownian particles that diffuse in two dimensions \cite{Chaikin,Mason,Mason2}. A vast body of experimental studies 
on these systems have clarified the importance of entropic particle interactions to stabilize different liquid-crystal
and solid phases with exotic symmetries beyond the standard uniaxial nematic (N) symmetry.
In particular, tetratic (T) and triatic (TR) liquid-crystal phases 
were found when the particles have cross sections with rectangular \cite{Chaikin}, or triangular \cite{Mason} 
geometry. Experiments with particles of square sections did not find the T phase \cite{Mason2}, but this
was due to the roundness of the corners, as proved by Monte Carlo (MC) simulations \cite{Escobedo}. 
The stability of T \cite{Schlacken,Frenkel,MR,Donev,Selinger,Sabi} and TR \cite{Dijkstra,MR2} phases as a function of 
particle shape was also confirmed by theoretical and simulation works on two-dimensional hard particle fluids.  
Experiments on monolayers of vibrated granular cylinders \cite{Menon1,Dani,GP1,GP2}
and squares \cite{Menon2} also showed the presence of stationary nonequilibrium T-like textures in the 
arrangement of particles. These results suggested that entropic interactions are also important to determine
the orientational ordering patterns observed in dissipative systems. Experiments
conducted on vibrated granular rods \cite{Ariel} and equilibrated colloidal silica rods \cite{Arts} 
under annular confinement showed the presence of topological defects and domains walls between regions
of different orientational and spatial ordering. This in turn suggests similarities
between dissipative and equilibrium systems in situations where entropic interactions play a dominant
role, i.e. at high packing fractions. Recently the T phase of kite-like particles was
also found \cite{Mason3,MR3}. How regular polygons order in liquid-crystal and crystalline phases
as density is varied strongly depends on the number of polygonal sides, an issue that was intensively
studied by MC simulations \cite{Glotzer}.

Recently Gantapara et al. \cite{Dijkstra} have conducted MC simulations on hard particles consisting of
equilateral and right isosceles
triangles. For the latter, 
the authors found the presence of a seemingly exotic liquid-crystal phase,
which they called Rhombic (R), between the isotropic (I) and its crystalline counterpart (the rhombic crystal). 
This phase seems to exhibit strong octatic (O) correlations, with a high value for corresponding order parameter, $Q_8$.
In an O phase the orientation distribution function, i.e. the probability density of a particle to
orient with respect to one of the equivalent directors at an angle $\phi$, has eightfold 
symmetry, $h(\phi)=h(\phi+\pi/4)$, in contrast with the T phase where the symmetry is $h(\phi)=h(\phi+\pi/2)$. 

\begin{figure*}
	\epsfig{file=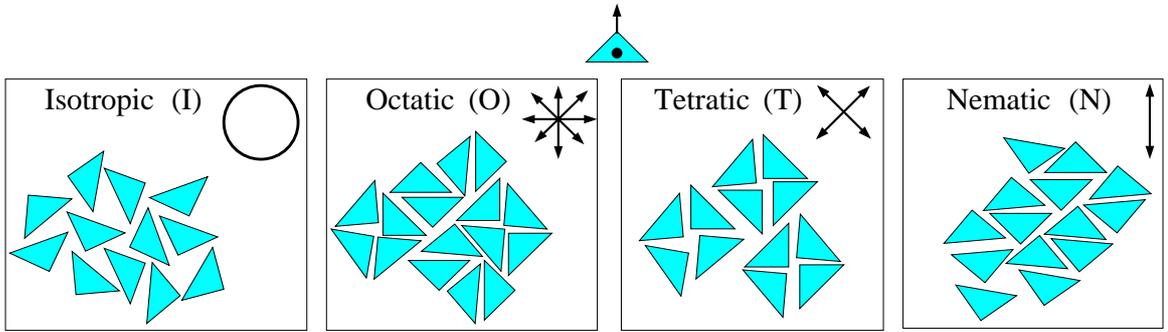,width=6.in}
	\caption{\label{fig0m}Sketches of different liquid-crystal phases of hard right 
	triangles: I, O, T and N, as labelled in the figure. The equivalent directors along which 
	particle axes of triangles (defined in the top) are aligned are also indicated.}
\end{figure*}

Fig. \ref{fig0m} shows a schematic of the I, O, T 
and N liquid-crystal phases of hard right triangles. One can anticipate the importance of 
particle clustering in this system, with particles easily forming square-like dimers or tetramers 
that build up O or T ordering, corresponding to eight or four equivalent directors, respectively.
Note that the axis of a right triangle is defined by a unit vector connecting the
barycenter and the right-angled vertex.

The phase discovered by Gantapara et al. is probably the standard T phase since,
as shown below, there are good reasons to expect that
the phase cannot be purely octatic. In any case, the main focus of the
authors of Ref. \cite{Dijkstra} was on the study of chirality in the crystal phase of hard equilateral and 
right-angled triangles.

We have recently applied a Density-Functional Theory (DFT) based on Scaled-Particle Theory (SPT)
to study the phase behavior of a fluid of hard isosceles triangles as a function of their opening angle \cite{MR2}. 
This theory is close in spirit and qualitatively similar in results to the standard Onsager theory \cite{Onsager} and to
extensions of Parsons-Lee-type theories \cite{Lee} applied to two dimensional particles.
For the particular case of right-angled triangles, we found that the only stable liquid-crystal phase predicted by
the theory was the standard uniaxial N phase, with the symmetry $h(\phi)=h(\phi+\pi)$.
This is in stark disagreement with the simulations of Gantapara et al. and points to a essential 
problem of the above theory and, by extension, of the (second-order virial) Onsager theory. 
We remind the reader that, by contrast, the SPT-based theory correctly predicts the existence of the exotic T phase in
fluids of hard rectangles with low aspect ratio and of the TR phase for a fluid of hard equilateral triangles \cite{Schlacken,MR,MR2}.

The failure of DFT to correctly describe, at least qualitatively, the thermodynamically
stable symmetry of an oriented fluid is unusual. It means that the important 
angular correlations that drive the system to the stable symmetry are not present at the level of
two-particle correlations but in higher-order terms. This failure brings to mind the inability of DFT to reproduce
the intermediate hexatic phase in systems of hard discs, although the origin of the problem is
completely different.
In this case even the most sophisticated, albeit still approximate,
versions of DFT applied to the hard disk fluid 
predict a first-order transition between the fluid and a crystal with particles at the nodes of 
a triangular lattice \cite{Roth}. However, simulations show an intermediate hexatic phase between
fluid and crystal exhibiting quasi-long-range bond-orientational ordering \cite{Strandburg,Mak,Engel},
associated to two-particle positional correlations. One can speculate on the possibility to construct
an accurate hard-disk density functional based on two-body density functions, as proposed for example 
in Ref. \cite{Iyetomi,Schiper}: coupling between first nearest-neighbor bonds and spatial coordinates 
would be contained explicitly in the theory by construction, and phases with bond orientational ordering
could in principle be stabilised.

In the present case, however, the order is orientational and characterised by a 
one-particle distribution function, namely the orientational distribution function $h(\phi)$, not
by a bond-orientational order. The origin of the failure is therefore different. Retrospectively,
it is remarkable how Onsager theory, based solely on two-particle correlations, i.e. on the lowest-order
virial expansion of the free energy, can describe oriented phases of fluids and fluid phase transitions.
Obviously this is because, for fairly elongated particles, the second-order term is by far the
dominant one in the virial expansion. It is therefore even more remarkable that, in the last few 
decades, Onsager theory and its extensions (e.g. SPT), all based on two-particle orientational correlations,
have been applied with success even in situations where the latter condition is clearly not validated, 
for example in 2D and/or in the case of anisotropic but not very elongated particles. 
It is plausible that the limits of validity of the theory may be revealed in some cases, 
and the fluid of hard right triangles is, to our knowledge, the first example of this breakdown.

Motivated by this negative finding, in the present article we report on the results
obtained using an extended SPT theory that goes beyond two-body correlations by incorporating 
three-body orientational correlations through the third virial coefficient. This is the obvious
step to take in an attempt to remedy the deficiencies of the standard theory.
From the lessons learned in the hard-rectangle case \cite{MR1}, it is known that higher-order correlations 
are important to improve the predictions on the stability of high-symmetry orientational 
phases, since they more correctly account for particle configurations typical of these phases. However, the
numerical implementation of such a theory is not easy: the third virial coefficient is
in fact a functional of the orientational distribution function, $h(\phi)$, with respect to which
the free energy has to be functionally minimised. In order to make calculations
feasible, we consider a projection of $h(\phi)$ on Fourier space, and represent the third-virial
functional by means of particular low-order moments that dominate the behavior in the neighbourhood of the 
bifurcation point (from the isotropic to the oriented phase). A combination of MC integration and
Gaussian quadratures is used to evaluate these moments with reasonable accuracy. Using these techniques,
a bifurcation analysis and the full minimization of the resulting DFT were implemented, which produced
valid results at least for packing fractions close to the bifurcation point. 

Interestingly, the results from the extended theory
are qualitatively similar to the predictions of the standard SPT-based theory:
the only orientationally-ordered stable phase is the uniaxial N phase, and the structure of the orientational
distribution function has no indication whatsoever of T or O correlations. The obvious conclusion is that
third-order correlations are still not the important ones in the stability of these phases. Due to the
difficulties associated with the implementation of even higher-order terms, the virial approach
suggested by Onsager theory in this system seems hopeless in practical terms.

To understand the problem in more depth, especially in connection to the stability of high-symmetry
orientational phases, we have also performed $NVT$-MC simulations of a fluid of 
hard right-angled triangles. In line with the findings of Gantapara et al. \cite{Dijkstra}, we found that,
under compression from the isotropic fluid, the system becomes ordered in configurations where 
particles exhibit strong O correlations. Under expansion from a perfect crystal 
formed by tetramers of triangles arranged in squares, Fig. \ref{clust}(a) (this is the stable crystal phase 
according to \cite{Dijkstra}),
we found a (possibly discontinuous) melting transition to a T phase (with fourfold symmetry and a high number
of tetramers). This T phase is stable (or metastable) for approximately the same range of packing fractions 
as that of the O phase obtained on compression,
and under further expansion transforms into an isotropic phase via a discontinuous
phase transition). All of these symmetries are confirmed from the behaviour of suitable order parameters that 
describe the T and O symmetries, and are visualised very directly by looking at the orientational distribution 
function $h(\phi)$. 

Although we have not traced out the stability boundaries of the I, O and T phases, 
these results definitively answer the question as to the inadequacy of a third-virial DFT to describe the 
orientational ordering symmetries of a fluid of right-angled triangles. Obviously the inclusion of 
even more high-order virial coefficients could improve the description of the fluid
phase behavior. However, as mentioned above,
the numerical implementation of such a theory would constitute a huge numerical 
task, much more demanding than standard MC simulations on reasonably sized systems. Clearly a radically
new approach is needed. One possibility is the formulation of new models based on the 
self-assembling of particles into clusters or super-particles of different shapes and sizes;
these clusters would in turn be oriented in such a way that the final orientation of monomers will
exhibit the new exotic symmetries. As recently shown \cite{Glotzer2}, the entropic hard-particle 
interactions enjoy some similarities with chemical-bonding interactions, which eventually give rise to the 
formation of clusters of particles or supra-molecular aggregates, respectively. This view could be 
fruitful in the present system and might represent a worthy activity for the future.

The article is organized as follows. Sec. \ref{B3} is devoted to introducing the theoretical model for
a fluid of hard right-angled triangles, a model based on the extended third-virial DFT. 
In Secs. \ref{bifurca} and \ref{minimin} we implement a bifurcation analysis and the full minimization of the model, 
which is valid for densities close to the bifurcation point. Also, in Sec. \ref{MonteCarlo}
we present MC simulations which confirm
the presence of orientational symmetries different from uniaxial. 
Finally some conclusions are drawn in Sec. \ref{conclusions}.


\section{Third-virial DFT}
\label{B3}

We have already pointed out that the SPT-based DFT (an effective second-order virial theory), fails to
predict the stability of the T and/or O liquid-crystal phases. These phases are characterised by
an orientational distribution function $h(\phi)$ with four- and eightfold symmetries, respectively:
$h(\phi)=h(\phi+n\pi/4)$ ($n=1,2$). By contrast, MC simulations \cite{Dijkstra} clearly point to the existence
of both these symmetries in the region between the isotropic and crystal phases, even though the nature of the
stable phase and the relative stability of the two phases still demand clarification (see below).

It is well known that the third and higher-order virial coefficients, $B_n$ with $n>2$, cannot be neglected 
if we are to obtain a quantitatively correct description of the phase behavior of hard elongated particles
in two dimensions \cite{Talbot,Rigby}. This can be explained in terms of the nonvanishing limit of the ratio $B_3/B_2^2$ 
as particle elongation goes to infinity (the so-called Onsager limit). Although the triangular geometry 
studied here cannot be defined in such terms, we expect the effect of the third virial coefficient to 
be sufficiently important as to merit its inclusion in a DFT approach. A possible theory was already 
proposed and applied by us in a study of the hard-rectangle-fluid \cite{MR1}, where the effect of the
third virial coefficient was found to be important (although not crucial to determine phase symmetry). 
The essential idea is to approximate the excess free-energy density functional per particle 
(in reduced thermal units) as
\begin{eqnarray}
	&&\varphi_{\rm exc}[h]=-\log(1-\eta)+\frac{\eta}{1-\eta}b_2[h]\nonumber\\
	&&+\left(\frac{\eta}{1-\eta}+\log(1-\eta)\right)\left(b_3[h]-2b_2[h]\right),
	\label{Eq1}
\end{eqnarray}
where $\eta=\rho a$ is the packing fraction, defined as the product of mean number density, 
$\rho$, and particle area of the right isosceles triangle, $a=l^2/2$, with $l$ the 
length of the equally-sized triangle sides. In Eqn. (\ref{Eq1}) the coefficients $b_k[h]$ are defined as
\begin{eqnarray}
	b_k[h]=\frac{B_k[h]}{a^{k-1}}-1,
	\label{lab}
	\end{eqnarray}
	where $B_k[h]$, $k=2$ and 3, are the second and third virial coefficients, manifestly
	functionals of the orientational distribution function $h(\phi)$:
	\begin{eqnarray}
		&&B_k[h]=\frac{1}{k}\left(\prod_{i=1}^k\int_0^{2\pi}d\phi_ih(\phi_i)\right) {\cal K}^{(k)}
		\left(\boldsymbol{\phi}\right), \nonumber\\
		&&\quad \boldsymbol{\phi}=\left(\phi_1,\cdots,\phi_k\right).
		\label{Eq5}
	\end{eqnarray}
	In turn, the kernels ${\cal K}^{(k)}(\boldsymbol{\phi})$ are spatial integrals of 
	products of Mayer functions $f({\bm r}_{ij},\phi_{ij})$ associated to particles $i$ 
	and $j$ with relative positions and orientations ${\bm r}_{ij}={\bm r}_j-{\bm r}_i$ 
	and $\phi_{ij}=\phi_j-\phi_i$, respectively:
	\begin{eqnarray}
		&&{\cal K}^{(2)}(\boldsymbol{\phi})=
		-\frac{1}{A}\left(\prod_{i=1}^2 \int_A d{\bm r}_i\right)  
		f({\bm r}_{12},\phi_{12})\nonumber\\
		&&=-\int_A d{\bm r} f({\bm r},\phi)
		=A_{\rm excl}(\phi), \label{Eq2}\\
		&&{\cal K}^{(3)}(\boldsymbol{\phi})=-\frac{1}{A}\left(\prod_{i=1}^3 
		\int_A d{\bm r}_i\right)  
		f({\bm r}_{12},\phi_{12})f({\bm r}_{23},\phi_{23})\nonumber\\
		&&\times f({\bm r}_{13},\phi_{13})\nonumber\\
		&&=-\int_A d{\bm r} \int_A d{\bm r}' f({\bm r},\phi)f({\bm r}',\phi')f({\bm r}'-{\bm r},\phi'-\phi),
		\label{Eq3}
	\end{eqnarray}
	with $A$ the total area. In the above expressions we have implemented the change of variables 
	${\bm r}\equiv {\bm r}_{12}$ and $\phi\equiv \phi_{12}$ (first and second integrals), and 
	${\bm r}'\equiv {\bm r}_{13}$ and $\phi'\equiv \phi_{13}$ (second integral).
	The object $A_{\rm excl}(\phi_{12})$ is the excluded area between two particles.  
	Using these expressions, Eqn. (\ref{Eq5}) for $k=2$ and 3 becomes 
	\begin{eqnarray}
		&&B_2[h]=\frac{1}{2}\int_0^{2\pi} d\phi \Psi_2(\phi) 
		{\cal K}^{(2)}(\phi), \\
		&&\Psi_2(\phi)=\int_0^{2\pi}d\phi_1 h(\phi_1)h(\phi_1+\phi), 
		\label{sub2}\\
		&&B_3[h]=\frac{1}{3}\int_0^{2\pi} d\phi\int_0^{2\pi} 
		d\phi'\Psi_3(\phi,\phi'){\cal K}^{(3)}(\phi,\phi'),\\ 
		&&\Psi_3(\phi,\phi')=\int_0^{2\pi}d\phi_1 h(\phi_1)
		h(\phi_1+\phi)h(\phi_1+\phi') \label{sub3}
	\end{eqnarray}
	The density expansion of Eqn. (\ref{Eq1})  gives
	$\displaystyle{\varphi_{\rm exc}[h]=B_2[h]\rho+\frac{1}{2}B_3[h]\rho^2}+\cdots$, 
	and consequently the pressure is $\beta p=\rho+\rho^2\partial \varphi_{\rm exc}/\partial\rho
	=\rho+B_2[h]\rho^2+B_3[h]\rho^3+\cdots$ (with $\beta$ the Boltzmann factor). The truncated
	expressions provide the exact low-density limit up to third order in density. 
	Thus our proposed theory treats two- and three-body correlations exactly.

	The ideal free-energy density functional per particle is, as usual, 
	\begin{eqnarray}
		\varphi_{\rm id}[h]=\log \eta-1+\int_0^{2\pi} d\phi h(\phi)\log{\left[2\pi h(\phi)\right]},
		\label{id}
	\end{eqnarray}
	where the thermal volume term has been dropped. Therefore our theory is completely defined as 
	$\varphi[h]=\varphi_{\rm id}[h]+\varphi_{\rm ex}[h]$.

	\section{Bifurcation analysis}
	\label{bifurca}

	In this section we perform a bifurcation analysis of the theory presented above.
	The bifurcation defines the instability of the I phase against 
	orientational fluctuations of some particular symmetry. We begin by considering 
	the first-order Fourier expansion of the $h(\phi)$ function:
	\begin{eqnarray}
		h(\phi)\simeq\frac{1}{2\pi}\left(1+h_n \cos{2n\phi}\right), 
	\end{eqnarray}
	with $h_n$ the first-order Fourier amplitudes. The indexes 
	$n=1,2,3,4$ account for uniaxial N, tetratic T, triatic TR, and octatic O symmetries, 
	respectively. Substituting this expression into Eqns. (\ref{sub2}) and (\ref{sub3}) gives, to
	lowest order:
	\begin{eqnarray}
		&&\Psi_2(\phi)\simeq \frac{1}{2\pi}\left(1+\frac{h_n^2}{2}\cos{2n\phi}\right), \label{psi2}\\
		&&\Psi_3(\phi,\phi')\simeq\frac{1}{(2\pi)^2}
		\left\{1+\frac{h_n^2}{2}
		\left[\cos{2n\phi}+\cos{2n\phi'}\right.\right.\nonumber\\
		&&\left.\left.+\cos{2n(\phi-\phi')}\right]\right\}.\label{psi3}
	\end{eqnarray}
	Consequently,
	\begin{eqnarray}
		&&B_2[h]\simeq \frac{1}{2}\left({\cal K}_0^{(2)}+\frac{1}{2}{\cal K}_n^{(2)}h_n^2
		\right),\label{insert1}\\
		&&B_3[h]\simeq \frac{1}{3}\left[{\cal K}_{00}^{(3)}+
		\left({\cal K}_{n0}^{(3)}+\frac{1}{2}{\cal K}_{nn}^{(3)}\right)h_n^2\right], 
		\label{insert2}
	\end{eqnarray}
where the following coefficients have been defined:
	\begin{eqnarray}
		&&{\cal K}_n^{(2)}=\frac{1}{2\pi}\int_0^{2\pi}d\phi {\cal K}^{(2)}(\phi)\cos(2n\phi),\nonumber\\
		&&{\cal K}_{nm}^{(3)}=\frac{1}{(2\pi)^2}\int_0^{2\pi}d\phi\int_0^{2\pi}d\phi' 
		{\cal K}^{(3)}(\phi,\phi')\nonumber\\
		&&\times\cos{2(n\phi-m\phi')}. \label{Eq6}
	\end{eqnarray}
	The Fourier expansion of the ideal free energy Eqn. (\ref{id}) gives 
	$\varphi_{\rm id}[h]\simeq \log{\eta}-1+h_n^2/4$. 
	Inserting Eqns. (\ref{insert1}) and (\ref{insert2}) into 
	(\ref{Eq1}), we obtain the expansion of the total free-energy 
	per particle as $\varphi=\varphi_{\rm I}+\Delta\varphi$, with 
	\begin{eqnarray}
		&&\varphi_{\rm I}=\log\left(\frac{\eta}{1-\eta}\right)-1+
		\left(\frac{{\cal K}_{00}^{(3)}}{3a^2}-1\right)\chi_1(\eta)\nonumber\\
		&&-\left(\frac{{\cal K}_0^{(2)}}{2a}-1\right)\chi_2(\eta),\label{la_I}\\
		&&\Delta\varphi=\left[\frac{1}{4}+
		\frac{1}{3a^2}\left({\cal K}_{n0}^{(3)}+\frac{1}{2}{\cal K}_{nn}^{(3)} 
		\label{brakets0}
		\right)\chi_1(\eta)\right.\nonumber\\
		&&\left.-\frac{{\cal K}_n^{(2)}}{4a}\chi_2(\eta)\right]h_n^2,
		\label{la_Delta}\\
		&&\chi_k(\eta)=\frac{\eta}{1-\eta}+k\log{(1-\eta)},\label{brakets}
	\end{eqnarray}
	with $\varphi_{\rm I}$ the I phase contribution. $\Delta\varphi$ is different from zero in the 
	presence of a weak orientational ordering.
	All that remains is to calculate the coefficients defined in Eqn. (\ref{Eq6}). As shown in \cite{MR2}, 
	the expression for ${\cal K}_n^{(2)}$, for the particular case of right-angled isosceles triangles, is
	\begin{eqnarray}
		&&\frac{{\cal K}_n^{(2)}}{2a}-\delta_{n0}=-\frac{1}{\pi(4n^2-1)}\nonumber\\
		&&\times \left[2+(-1)^n+
		2\sqrt{2}\cos\left(\frac{n\pi}{2}\right)\right],
	\end{eqnarray}
	with $\delta_{ij}$ the Kronecker-delta.
	The coefficients ${\cal K}_{nm}^{(3)}$ cannot be calculated analytically and have been obtained numerically. MC integration was used to evaluate 
	the spatial kernels in (\ref{Eq3}), while the angular integrals in Eqn. (\ref{Eq6}) were evaluated
	using Gaussian-Legendre quadratures.
	$2\times 10^6$ random configurations of three particles with fixed
	orientations were generated, using the usual protocol: one particle is located at the origin, two
	particles are generated at random, both overlapping with the first one,
	thus ensuring that the first two Mayer functions are equal to $-1$, and the third Mayer function
	connecting the latter is checked. In the case of the angular integrals, a density of
	12 points per period were used for the four cases $n=1,\cdots,4$. The estimated error 
	of the coefficients is less than $0.1\%$.

	The minimization of the expanded total free-energy 
	per particle $\varphi[h]$ with respect to $h_n$ gives the bifurcation condition 
	$\partial \Delta\varphi/\partial h_n=0$ for $h_n\neq 0$, with $\Delta\varphi$ given by (\ref{brakets0}). 
	The solutions to this equation are obviously the same as the zeros of the 
	term enclosed in brackets.
	We have solved numerically this equation for the values of packing fractions 
	$\eta_n$ at bifurcation, considering separately the cases $n=1,2,3$ and $4$.  
	The results are collected in Table \ref{tabla}. Also included are the results 
	obtained from the SPT approach, which can be obtained easily by considering the 
	first two terms of Eqn. (\ref{Eq1}) for the excess free-energy per particle.
	In this case the bifurcation points are given analytically by 
	\begin{eqnarray}
		\eta_n^{(\rm SPT)}=\frac{1}{1-{\cal K}_n^{(2)}/a}.
		\label{spin}
	\end{eqnarray}

	\begin{table}
		\begin{tabular}{||c|c|c|c|c||}
			\hline\hline
			$n$ & 1 & 2 & 3 & 4\\ \hline 
			\ $\eta_n^{(\rm SPT)}$ \ & \ 0.8249 \ & \ $0.9928^*$ \ & \ 0.9821 \ & \ 0.9444 \ \\ 
			\hline
			\ $\eta_n^{(\rm B_3)}$ \ & \ 0.7325 \ & \ $0.9794^*$ \ & 0.9328 \ & \ 0.8353 \ \\ \hline 
			\ $\Delta \eta^{(\rm SPT)}_n$ \ & \ -- \ 
			& \ 0.1679 \ & \ 0.1572 \ & \ 0.1195 \  \\ \hline
			\ $\Delta \eta^{(\rm B_3)}_n$ \ & \ -- \ &\ 0.2469 \ &\ 0.2003 \ & \ 0.1062 \ \\  
			\hline\hline
		\end{tabular}
		\caption{The bifurcated values of packing fractions $\eta_n^{(\alpha)}$ (with $\alpha=$ SPT or 
		$B_3$) resulting from SPT and $B_3$ approaches corresponding to bifurcations from I 
		to N ($n=1)$, T ($n=2$), TR ($n=3$) two numbers means that 
		O ($n=4$) phases. The differences 
		$\Delta \eta_n^{(\alpha)}\equiv \eta_n^{(\alpha)}-\eta_1^{(\alpha)}$ 
		for $n\geq 2$ are also shown. An asterisk
		indicates that the value does not correspond to a true bifurcation at a
		second-order transition; this is because a first-order I-T metastable transition
		is found using the SPT theory and a parameterized $h(\phi)$ containing only 
		one parameter. The same behavior is expected for the $B_3$ approach.}
		\label{tabla}
	\end{table}

	A first conclusion that can be drawn from the results of the table 
	is that, compared to the second-virial approximation, the third-virial approximation 
	dramatically decreases the bifurcation packing fractions from the I phase to the orientationaly ordered 
	phases. In general the third-virial approximation gives a value for the I-N bifurcation,
	$\eta_1^{(\rm B_3)}\approx 0.73$, which is quantitatively similar 
	to that for the I phase coexisting with
	the liquid-crystal phase, as predicted by simulations \cite{Dijkstra}.
	Note again that simulations predict a very different symmetry (four- or eight- instead of two-fold) 
	for the latter phase: both, second- and third-virial approaches, predict a stable uniaxial N phase
	beyond the I phase.

	The failure of DFT to predict the correct symmetry for the liquid-crystal phase could have one or
	two of the following explanations: (i) the importance of four and higher-order
	particle correlations, quantified through the corresponding virial coefficients,
	to correctly predict the stability of the T/O phase
	(note that, strictly speaking, the theory generates these virial coefficients from a 
	density expansion; however, the angular correlations will be poorly represented);
	(ii) the presence of strong clustering effects,
	which a low-order virial theory would be unable to account for.
	In any case, the sequence 
	$\eta_1^{(\alpha)}<\eta_4^{(\alpha)}<\eta_3^{(\alpha)}$ (see Table \ref{tabla}),
	together with the fact that from a value $\eta_2^*\approx 0.955$ a 
	(highly ordered) metastable T solution is found
        (see discussion in Section \ref{minimin}), and taking into account the 
	corresponding differences 
	$\Delta \eta_n^{(\alpha)}\equiv \eta_n^{(\alpha)}-\eta_1^{(\alpha)}$ and 
	$\Delta\eta_2^*=\eta_2^*-\eta_1^{(\rm spt)}\approx 0.13$, 
	we can conclude that, within the context of DFT theory, the TR,  T and O 
	symmetries 
	can be discarded as possible 
	liquid-crystal symmetries for hard right isosceles triangles.

	\section{Minimization}
	\label{minimin}

	A next step beyond the bifurcation analysis consists of analysing the equilibrium solutions
	of the extended density functional by minimization. 
	An important motivation for this calculation is to examine the orientational
	distributions and search for the possible existence of secondary peaks in
	$h(\phi)$ at $\phi=\{\pi/4,\pi/2,3\pi/4\}$ in the interval $\phi\in[0,\pi]$
	(which would point to four- or eightfold orientational symmetry).
	Then, we are probing the ability of three-body correlations to capture this symmetry.
	First we introduce a parameterization
	for the orientational distribution function, which we simply take as a truncated Fourier expansion:
	\begin{eqnarray}
		h(\phi)=\frac{1}{2\pi}\left(1+\sum_{k=1}^{n_{\rm max}} h_k\cos(2k\phi)\right),
		\label{expansion}
	\end{eqnarray}
	We choose $n_{\rm max}=5$, which will be valid sufficient close to the 
	bifurcation point. 
	Substituting (\ref{expansion}) into Eqns. (\ref{psi2}) and (\ref{psi3}), we obtain 
	\begin{eqnarray}
		&&\Psi_2(\phi)=\frac{1}{2\pi}\left(1+\frac{1}{2}\sum_{k=1}^{n_{\rm max}}h_k^2\cos(2k\phi)\right),\\
		&&\Psi_3(\phi,\phi')=\frac{1}{(2\pi)^2}\left\{1+\frac{1}{4}
		\sum_{(k_1,k_2)\neq(0,0)}^{\text{max}(k_1+k_2)=
		n_{\rm max}} h_{k_1}h_{k_2}
		\nonumber\right.\\
		&&\left. \times h_{k_1+k_2}\left[\cos(2(k_1\phi+k_2\phi'))\right.\right.\nonumber\\
		&&\left.\left.+\cos(2(k_2\phi-(k_1+k_2)\phi'))
		\right.\right.\nonumber\\
		&&\left.\left.+\cos(2(k_2\phi'-(k_1+k_2)\phi))\right]\right\}.
	\end{eqnarray}
	After substitution into Eqns. (\ref{Eq1}) and (\ref{id}), the total free energy per particle
	$\varphi=\varphi_{\rm I} +\Delta \varphi$ is obtained, with $\varphi_{\rm I}$ calculated from (\ref{la_I}).
	The excess part over the isotropic contribution, $\Delta \varphi$, is now
	\begin{eqnarray}
		&&\Delta \varphi=\int_0^{2\pi} d\phi h(\phi) \log {\left[2\pi h(\phi)\right]}+
		\frac{1}{6a^2}\sum_{(n,m)\neq (0,0)}^{n+m\le n_{\rm max}}
		h_{n}h_{m}\nonumber\\
		&&\times h_{n+m} \left[
		      \frac{{\cal K}^{(3)}_{-n,m}}{2}+{\cal K}^{(3)}_{n,n+m}\right]\chi_1(\eta)\nonumber\\
	      &&-\frac{1}{4a}\left(\sum_{n\neq 0}^{n_{\rm max}} h_n^2{\cal K}_n^{(2)}\right)\chi_2(\eta).
		\label{double}
	\end{eqnarray}
	It is understood that $h_{n}=1$ for $n=0$ in the double sum of Eqn. (\ref{double}).
	As in Section \ref{bifurca} the coefficients ${\cal K}^{(3)}_{nm}$ are evaluated numerically
	using MC integration and Gaussian quadrature (note that in this case a total of 71 coefficients 
	need to be evaluated instead of only 9 in the bifurcation analysis).
	The total free energy is then minimized with respect to the Fourier amplitudes 
	$\{h_k\}$ ($k=1,\cdots,n_{\rm max}$) ($n_{\rm max}=5$) using a Newton-Raphson procedure.

	\begin{figure*}
		\hspace*{0.8cm}
		\includegraphics[width=7.in]{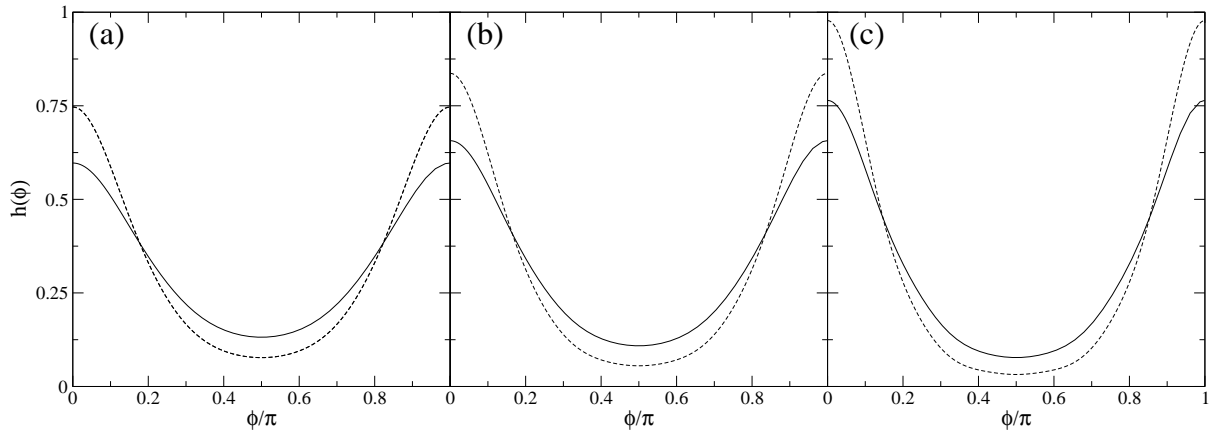}
	\caption{Distribution functions $h(\phi)$ of a N fluid of hard right triangles obtained from the 
		minimization of: (i) SPT theory (dashed curve), (ii) Extended SPT theory (solid curve).
		Three cases of different densities are shown. (a) $\eta=0.8437$ (SPT) and $0.7500$ (Extended SPT).
		(b) $\eta=0.8500$ (SPT) and $0.7547$ (Extended SPT).
		(c) $\eta=0.8600$ (SPT) and $0.7636$ (Extended SPT). In each case
		densities have been chosen with the same relative departure from the respective
		bifurcation densities: $3.0\%$ in (a), $4.2\%$ in (b), and $5.5\%$ y (c). In panel (a)
		SPT results are plotted for the cases with $n_{\rm max}=5$ and $n_{\rm max}=40$ Fourier 
		coefficients; both results cannot be distinguished at the scale of the graph. The extended SPT theory
		has been calculated with $n_{\rm max}=5$ Fourier coefficients, as explained in the text.}
		\label{fig0}
	\end{figure*}

Fig. \ref{fig0} shows an example of a minimization of the $B_3$-extended SPT approach 
for a packing fraction $\eta=0.75$;
	this is not far from the I-N bifurcation point at $\eta_1=0.7325$.
The orientational distribution function $h(\phi)$ is represented, together with the result from
	the second-virial SPT theory of Ref. \cite{MR2}. In the latter case $h(\phi)$ was calculated using
	$n_{\rm max}=5$ and also $n_{\rm max}=40$, which practically gives the exact result considering 
	that the last Fourier coefficient, $h_{40}$, obtained from the minimization  
	has an absolute value of less than $10^{-7}$. The functions of panel (a) (dashed curve),
	cannot be distinguished one form the other at the scale of the graph, their mean square
difference being $\epsilon\equiv \sqrt{\int_0^{2\pi}d\phi \left(h(\phi)-\tilde{h}(\phi)\right)^2}=
		5.2\times 10^{-4}$.  The packing fraction $\eta=0.8437$ was chosen
	so as to give the same relative distance $(\eta-\eta_1)/\eta_1$ from the I-N bifurcation point
as the $B_3$-extended theory (solid curve). We can see that the two theories give a qualitatively similar
orientational structure, with the function from the extended theory being more weakly oscillatory because
of the smaller density. We also show the functions obtained from the SPT and $B_3$-extended theories 
for higher packing fractions corresponding to the 
same relative departures from the respective bifurcation densities.
What is important from these calculations is that neither theory predicts 
any trace of secondary peaks and, consequently, of tetratic or octatic correlations, in contrast with the
simulation results.

\begin{figure}
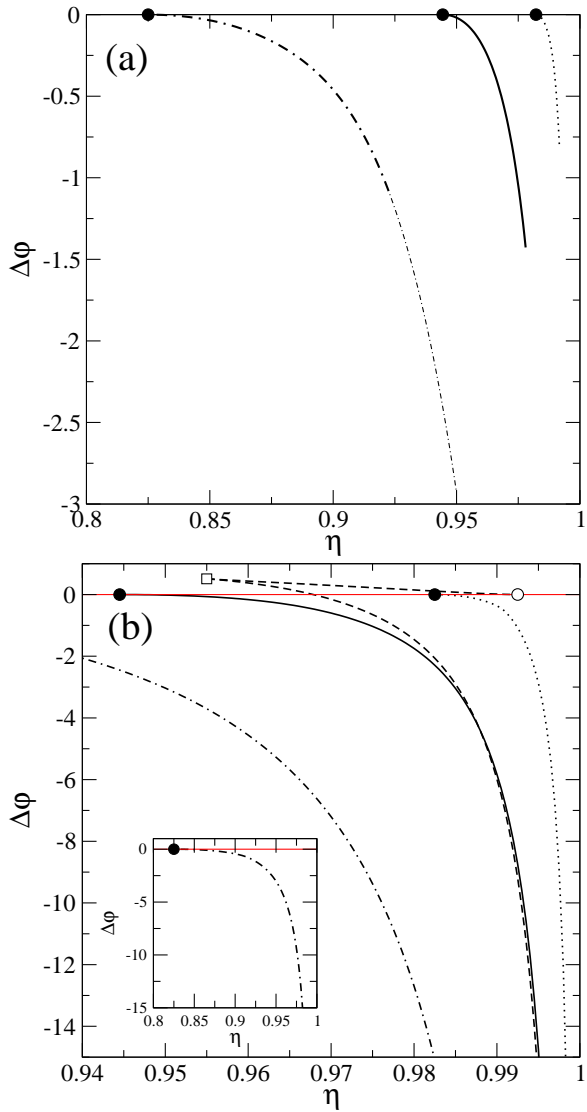

			\epsfig{file=fig3a.eps,width=3.in}
			\epsfig{file=fig3b.eps,width=3.in}
			\caption{Free-energy differences $\Delta\varphi\equiv \varphi_{\alpha}
			-\varphi_{I}$, with $\alpha$= N (dot-dashed), O (solid), T (dashed) and 
			TR (dotted), as a function of packing fraction $\eta$, using the 
			Fourier-coefficient (a) and one-parameter (b) minimizations. 
			In (a) we show with a tiny dot-dashed line a simple extrapolation 
			of the N branch to higher densities. The inset 
			in (b) show the complete N branch. The filled circles in (a) and (b) show 
			the bifurcation points at second order transitions while the open 
			one in (b) labels its first-order counterpart in the T branch. The 
			open square show the location of the first metastable solution with 
			T symmetry.}
			\label{fig_nueva}
		\end{figure}

One may wonder which of the following scenarios takes place at densities above 
the I-O bifurcation: (i) The O and N free-energy branches cross each other at some density, or (ii) 
the N branch continues to be the lowest one. In order to investigate this point, we minimised the
SPT functional restricting the set of Fourier coefficients $\{h_k\}$ to only those with index $k=4j$
(giving a distribution $h(\phi)$ with perfect O symmetry). A free-energy branch was generated
for a density interval starting at the I-O bifurcation point and up 
to densities where the octatic order parameter is $Q_8\simeq 0.97$ (in this situation
the truncated Fourier series still gives correct results). 
We also calculated the N branch up to densities for which $Q_2\simeq 0.97$.
As expected, we obtained a ``metastable'' O phase.  However, the first scenario above can be
discarded, as the difference between the O and N free-energy branches is huge 
(note that the latter
bifurcates at much lower densities); this conclusion results from a simple
                extrapolation to the region 
		where the O phase is metastable. The situation is
		even worse in the case of the metastable TR phase, as it bifurcates at higher 
		packing fraction 
		(see the table \ref{tabla}). All these results are shown 
		in Fig. \ref{fig_nueva}(a), where we plot the free-energy 
		differences  
		$\Delta\varphi\equiv \varphi_{\rm \alpha}-\varphi_{\rm I}$ 
		($\alpha=$N,O,TR) between $\alpha$ and I phases
		calculated from their respective 
		bifurcation points. 
		We remark that the T phase bifurcates 
		from the I phase but within a first-order transition scenario 
		(see comment in Table \ref{tabla}), unlike the N, O and TR phases. 
		This point is elucidated not via Fourier-amplitude minimization 
		(we were unable to obtain a metastable T solution with the proper 
		restrictions over $\{h_k\}$), but from a simple one-parameter 
		($\lambda$) minimization of a parameterized orientation distribution function,
		\begin{equation}
			h(\phi)=\dfrac{e^{\lambda \cos(2n\phi)}}{\pi I_0(\lambda)},\hspace{0.6cm}
		\phi\in[0,\pi],
	        \end{equation}
		with $I_0(x)$ the zeroth-order modified Bessel function of the first kind.
		The results are shown in Fig. \ref{fig_nueva}(b):
		from the bifurcation point (open circle in the figure), 
		located at a packing fraction $\eta=0.9928$  
		(see Table \ref{tabla}), an unstable T branch departs towards lower densities,
		with a free energy higher than that of the I phase. This branch
		terminates at $\eta\approx 0.955$ (open square in the figure), where the T phase
		becomes metastable for the first time, with a high order parameter $Q_4$. 
		From this point a second T branch develops towards higher densities,
		$\Delta\varphi$ eventually becoming negative at $\eta\approx 0.968$, 
		indicating that the T phase is more stable than the I phase. 
		This is the usual scenario for a first-order phase transition.
		When $\eta$ further increases from this value, the T free-energy branch 
		also crosses the O branch at $\eta\approx 0.988$.  
		In any case, the free energy of the N phase has by far the lowest free energy, 
		as can be seen in Fig. \ref{fig_nueva}. The crossing of the T and O branches
		at high packing fractions is interesting. It can be understood by invoking
		the limit value of the scaled second-virial coefficient (\ref{lab}) as 
		$\lambda\to\infty$:  
		\begin{eqnarray}
			&&b_2^{(n)}\equiv \lim_{\lambda\to\infty} b_2[h]\nonumber\\
			&&=\frac{1}{2an}\left[\frac{A_{\rm excl}(0)+A_{\rm excl}(\pi)}{2}
			+\sum_{k=1}^{n-1}A_{\rm excl}\left(\frac{k\pi}{n}\right)\right]-1.
			\nonumber\\
		\end{eqnarray}
		Inserting the known analytic expression for $A_{\rm excl}(\phi)$, we obtain 
		$b_2^{(1)}<b_2^{(2)}\lesssim b_2^{(4)}<b_2^{(3)}$ for the
		N ($n=1$), T ($n=2$), TR ($n=3$) and O ($n=4$) symmetries. This fact explains 
		the reason for the crossing behavior: the double-averaged excluded area with 
		respect to $h(\phi)$ for the T symmetry, although similar in magnitude,
		is lower than that obtained for the O symmetry. Obviously this occurs only
		at very high densities, when the orientational order is almost perfect 
		and the above asymptotic expression can be justified. 
		The fact that the O-phase free-energy is lower than that of the T phase
		at lower densities implies that the opposite behavior is true when the 
		orientational distribution function is less sharply peaked. 

                Even though these calculations were 
restricted to the SPT theory, we can be quite confident that the similar scenario results from
the $B_3$ theory (note that the latter theory is difficult to implement at high densities 
due to the prohibitively large number of three-body kernels 
projected on Fourier space that are necessary to describe a phase with high orientational order).

\begin{figure}
\epsfig{file=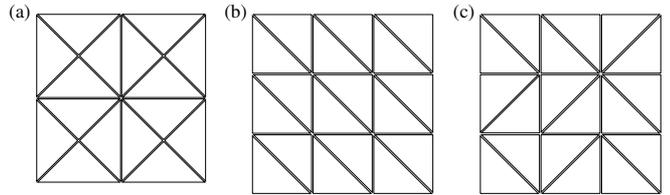,width=3.5in}
	\caption{\label{conf} 
	Possible particle configurations for crystal phases of right triangles. 
	(a) Square lattice of tetramers. Each tetramer is formed by four triangles. This is
	the starting configuration of expansion run indicated by open circles in Fig. \ref{ops}(a).
	(b) Square lattice of dimers with ordered orientations. This is the starting configuration
	of expansion run shown in Fig. \ref{ops2}. (c) Square lattice of dimers with disordered 
	orientations. This is the starting configuration of expansion run shown in Fig. \ref{ops1}.}
\end{figure}

\section{Monte Carlo simulations}
\label{MonteCarlo}

Gantapara et al. \cite{Dijkstra} have presented a detailed Monte Carlo simulation 
study of the fluid and crystal phases of hard right-angled triangles.
Their isothermal-isobaric simulations on systems of 1600 particles point to the existence of 
a liquid-crystal phase between the isotropic fluid and the crystal. Based on the analysis
of order parameters and angular correlation functions, this phase was named 
Rhombic (R) and characterised by the $Q_8$ order parameter. Order parameters are defined, in terms of the orientational
distribution function, as
\begin{eqnarray}
	Q_k=\int_0^{2\pi}d\phi h(\phi)\cos{(k\phi)}.
\end{eqnarray}
These order parameters characterise the N ($k=2$), T ($k=4$), TR ($k=6$) and O ($k=8$) symmetries.
The term `rhombic' used in \cite{Dijkstra} reflects
the structure of the stable crystal phase, Fig. \ref{conf}(a), which consists of a square lattice of tetramers, 
each formed by four triangles in a square configuration \cite{Dijkstra}. Unfortunately, the information
provided by Gantapara et al. does not allow us to ascertain the true symmetry of the liquid-crystal phase.
A pure octatic phase would be characterised by 
$Q_8>0$ and $Q_2=Q_4=0$, and by an orientational distribution function $h(\phi)$ 
with eight equivalent (equal-height) peaks in the interval $[0,2\pi)$, fulfilling
the condition $h(\phi)=h(\phi+\pi/4)$. A focus on the $Q_8$ order parameter
in \cite{Dijkstra} may indicate the occurrence of octatic symmetry in the 
intermediate phase between isotropic and crystal phases. 

To further investigate 
this issue in more detail, and to obtain a test bed for the DFT results, we performed 
$NVT$ Monte Carlo simulations on samples of 576 triangles (some selected $NpT$ simulations were 
also performed to investigate some specific issues). Although our sample sizes are smaller, 
our results complement the work of Gantapara et al. in the sense that they focus on the complete
set of order parameters (rather than only on $Q_8$), and also on the orientational
distribution function. 

\begin{figure*}
\epsfig{file=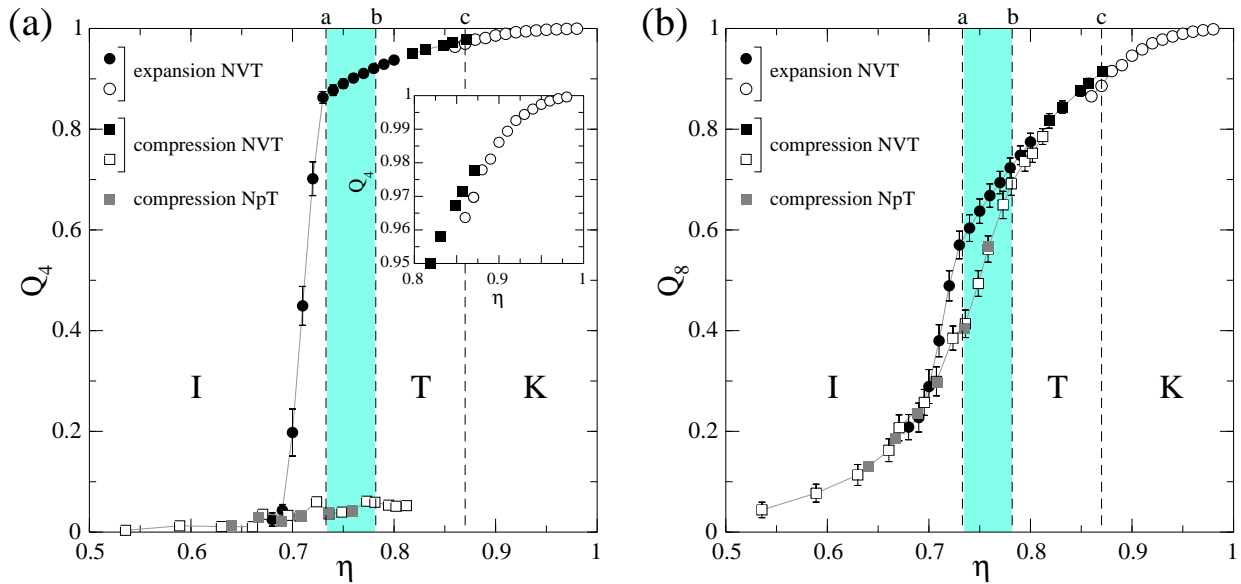,width=6.5in}
	\caption{\label{ops} 
	Order parameters $Q_k$ (with $k=4$ and $8$) as a function of packing fraction $\eta$, as
	obtained from $NVT$ and $NpT$ Monte Carlo simulations on 576 hard right triangles.
	Compression runs are represented by squares, while expansion runs are represented with
	circles. Panel (a) shows the $Q_4$ order parameter, while (b) shows the $Q_8$
	order parameter. In both panels the different runs, indicated in the key, are as follows:
	filled circles, NVT expansion from an initial crystalline configuration of tetramers 
	at $\eta=0.8$ shown in Fig. \ref{conf}(a); 
	open circles, NVT expansion from an initial crystalline configuration
	of tetramers at $\eta=0.98$; filled squares, NVT compression from an initial
        crystalline configuration of tetramers at $\eta=0.82$; open squares, NVT compression
        from an initial isotropic configuration at $\eta=0.54$; gray squares, NpT
	compression from an initial isotropic configuration at low pressure.
	The isotropic-liquid-crystal coexistence reported by \cite{Dijkstra} is shown
	as a shaded region, bounded by vertical dashed lines corresponding to the
	transition densities: a, coexistence density for isotropic
	at $0.733$. b, coexistence density for liquid-crystal phase at $0.782$. Also, the vertical line
	c is the density of the transition from the 
	liquid-crystal phase to the crystal phase of tetramers at $0.87$ obtained in \cite{Dijkstra}. 
	The order parameters $Q_2$ and $Q_6$
	are not represented as their values are $<0.1$ for all densities and runs. 
	Nonvanishing values of $Q_8$ below $\eta=0.733$ in panel (b) might be a finite-size effect.
	Regions of stability of the I, T and K phases, according to \cite{Dijkstra}, 
	are indicated with the corresponding labels. 
	The inset in panel (a) is a zoom of the liquid-crystal--crystal phase transition.}
\end{figure*}

\begin{figure}
\epsfig{file=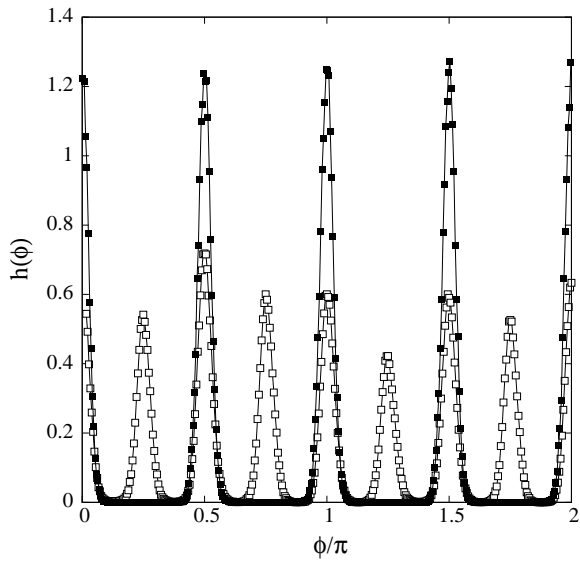,width=3.in,angle=-90}
	\caption{\label{hphi} 
	Orientational distribution function $h(\phi)$ from simulation, for two cases: $\eta=0.802$,
	obtained by compression from the isotropic fluid (open squares); and $\eta=0.8$, 
	obtained by equilibration of a perfectly ordered configuration of tetramers (filled squares).}
\end{figure}

Several $NVT$ runs were performed, following compression or expansion starting from
configurations with different symmetries. All runs comprised $2\times 10^5$ MC steps for equilibration and
$3\times 10^5$ MC steps for averaging. $NpT$ runs were $10$ times longer.
Figs. \ref{ops} shows the order parameters $Q_k$, with $k=4$ and $8$, for some of the runs.
Results for compression and expansion runs are shown by squares and circles,
respectively. Two compression runs in the NVT ensemble were performed. The first (open squares)
started at a low density, $\eta=0.536$, in the isotropic fluid region, and continued up to a 
density $\eta=0.812$. Along this run all order parameters are low except $Q_8$, 
which shows a steady increase with density. This seems compatible with the results of Gantapara et al.
Since their system size is three
times larger, changes in the order parameter near the isotropic-liquid-crystal
transition are more abrupt in their case. At the end of our compression run values of the order 
parameters are $Q_2=0.023\pm 0.003$, $Q_4=0.053\pm 0.004$, $Q_6=0.023\pm 0.006$ and 
$Q_8=0.785\pm 0.016$, pointing to octatic symmetry.
The orientation distribution function for a density $\eta=0.802$ is plotted in Fig. 
\ref{hphi}, using open squares. The peak heights are not exactly the same, which
may due to statistical fluctuations or imperfect orientational sampling.
To check the compression $NVT$ results, additional $NpT$ simulations were performed (grey symbols
in Fig. \ref{ops}). These results confirm the previous findings on the octatic-like phase
obtained by compressing the system from the isotropic phase.

However, the identification of the liquid-crystal phase as an octatic phase is challenged by the 
results from our expansion runs. These are indicated in Fig. \ref{ops} 
by means of circles. We focus first on the run starting at $\eta=0.8$ (filled circles) 
in a (crystalline) configuration of tetramers, Fig. \ref{conf}(a). This density was chosen because
the free energy calculations reported in Ref. \cite{Dijkstra} led the authors to 
conclude that the isotropic and liquid-crystal phase coexist along the
density gap $0.733$--$0.782$, indicated in Fig. \ref{ops} by a shaded region bounded
by vertical lines labelled `a' and `b', respectively. 
A quasiperfect arrangement of tetramers would give $Q_4\simeq 1\agt Q_8$.
From this perspective, it is no surprise that the largest order parameter in this expansion run is
$Q_4$. The system finally collapses to the isotropic phase below $\eta\simeq 0.73$.
The orientational distribution function along this run (see Fig. \ref{hphi})
confirms that the equilibrium configuration corresponds to a T phase, with 
four equivalent peaks in the interval $[0,2\pi)$, fulfilling the condition $h(\phi)=h(\phi+\pi/2)$.
Note that no signs of O symmetry, in the form of local maxima at $\pi/4$, $3\pi/4$, $5\pi/4$ and
$7\pi/4$, are visible. We conclude that the T phase is at least metastable for the 
system size used, but our analysis cannot say anything definite as to the true stable phase in
the interval $0.733<\eta<0.782$, either O or T
(detailed free-energy calculations would be necessary to settle this question). 

\begin{figure}
\epsfig{file=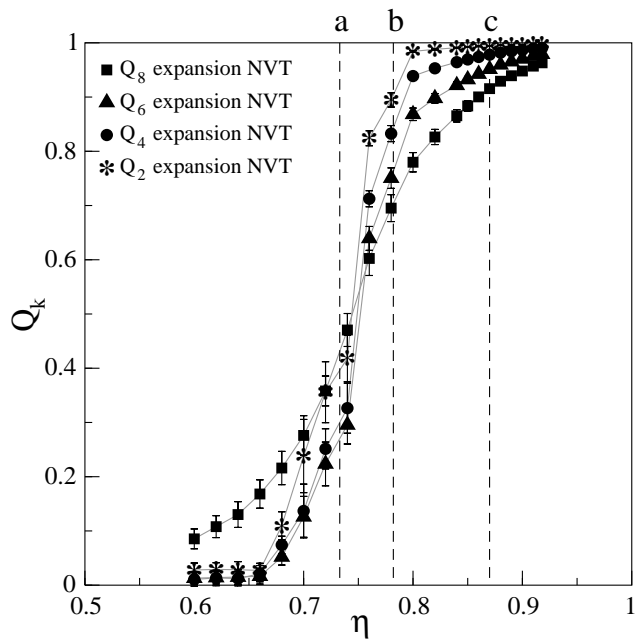,width=3.5in}
	\caption{\label{ops2} 
	Order parameters $Q_k$ (with $k=2, 4, 6$ and $8$) as a function of packing fraction $\eta$, as
	obtained from $NVT$ Monte Carlo simulations on 576 hard right triangles. A single expansion run
	from an initial configuration of ordered dimers 
[shown in Fig. \ref{conf}(b)], at $\eta=0.92$ is shown.
	Meanings of symbols are explained in the keybox. Vertical dashed lines correspond to the
        transition densities reported in \cite{Dijkstra}: a, coexistence density for isotropic.
        b, coexistence density for liquid-crystal phase. c: density of the transition from the
	liquid-crystal phase to the crystal phase of tetramers. }
\end{figure}

Compression and expansion runs were also conducted to study the 
transition between liquid-crystal and crystal phases. In the compression run (filled squares)
we started from the equilibrated point at $\eta=0.8$ of the previous
expansion run (filled circles). The expansion run (open circles) starts
from a perfect crystal of tetramers, Fig. \ref{conf}(a), at $\eta=0.98$ which, as shown
by the simulations of Gantapara et al. \cite{Dijkstra}, is the most stable crystal phase). 
The path ends at a density of $\eta=0.86$. These runs indicate
the presence of a first-order phase transition between T and K (crystal) phases, 
since the two branches are
not connected smoothly, see inset (a vertical line labelled `c' in Fig. \ref{ops} at density
$\eta=0.87$ indicates the density of the reportedly continuous transition \cite{Dijkstra}). 

It is difficult to explain the existence of a strict octatic symmetry for the 
liquid-crystal phase of hard right triangles on purely theoretical grounds. 
The reason is that this symmetry requires a delicate balance of particle arrangements so as to give 
equivalent orientations at multiples of $\pi/4$. We are inclined to think that the global
symmetry of the stable phase should be tetratic, 
but the problem persists that simulations indicate two competing liquid-crystal phases,
one phase possessing strong octatic correlations and another phase without such correlations.

In any case, the results reported up to this point indicate that the liquid-crystal phase of hard right triangles is not the
standard uniaxial N phase, as predicted by DFT, 
but rather a more symmetric phase with tetratic symmetry and possibly strong octatic correlations. 
In order to investigate this result from the point of view of simulation, we performed additional simulations,
following two strategies. First, we conducted expansion runs from the crystal region, starting with particle 
configurations different from a square lattice of tetramers. We stress that, according to the
free-energy calculations of Gantapara et al. \cite{Dijkstra},
the crystal of tetramers, Fig. \ref{conf}(a), is slightly more stable than
the crystal of dimers, Fig. \ref{conf}(b). The latter has uniaxial symmetry and can be considered as the crystalline `precursor'
of the uniaxial liquid-crystalline phase predicted by DFT. An obvious question is: could the uniaxial N
phase be at least metastable with respect to the other, more plausible symmetries? Fig. \ref{ops2} presents 
an expansion run from a crystal phase made of dimers with identical orientations at density
$\eta=0.92$. The crystal transforms 
continuously into a fluid phase with the same uniaxial symmetry (all order parameters are high and in the
order $Q_2>Q_4>Q_6>Q_8$). Along the simulations no indication of even weak tetratic or octatic correlations
could be detected: the orientational distribution function $h(\phi)$ (not shown) only exhibits
peaks at $0$ and $\pi$, with local maxima at other angles being completely absent even at 
low densities. This is an indication that the uniaxial
N phase predicted by DFT may be metastable with respect to other more symmetric phases, at
least at the scale of our $10^6$ MC-step simulations.

\begin{figure}
\epsfig{file=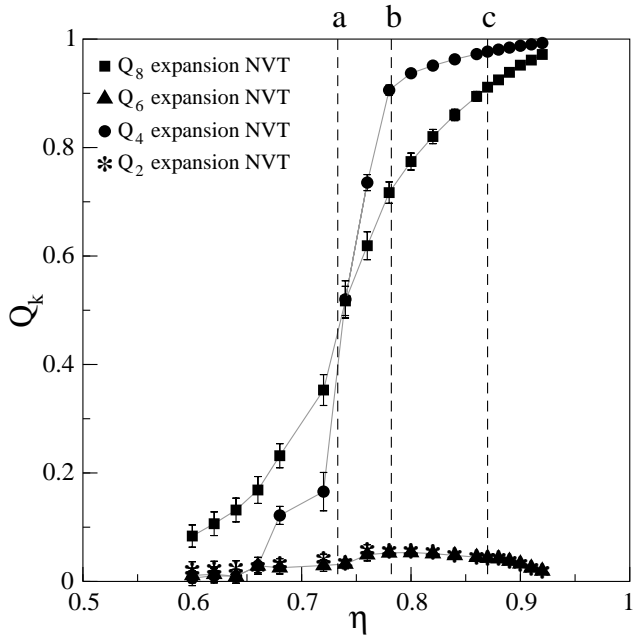,width=3.5in}
	\caption{\label{ops1} 
	Order parameters $Q_k$ (with $k=2, 4, 6$ and $8$) as a function of packing fraction $\eta$, as
	obtained from $NVT$ Monte Carlo simulations on 576 hard right triangles. A single expansion run
	from an initial configuration of disordered dimers 
	[shown in Fig. \ref{conf}(c)] at $\eta=0.92$ is shown.
	Meanings of symbols are explained in the keybox.
	Vertical dashed lines correspond to the
        transition densities reported in \cite{Dijkstra}: a, coexistence density for isotropic.
        b, coexistence density for liquid-crystal phase. c: density of the transition from the
	liquid-crystal phase to the crystal phase of tetramers. }
\end{figure}

A second run was performed on the same system but now from a configuration consisting 
of a square lattice of dimers with random orientations, Fig. \ref{conf}(c). The results are shown 
in Fig. \ref{ops1}.
This configuration was not considered by Gantapara et al. as a
possible candidate for equilibrium crystal phase, but clearly enjoys a higher entropy than the 
uniaxial
crystal of dimers and could compete with the crystal of tetramers (note that the entropy difference per
particle between the crystal of tetramers, Fig. \ref{conf}(a), and the crystal of orientationally 
ordered dimers, Fig. \ref{conf}(b), is only $0.011k$ per particle at $\eta=0.91$, the only density
investigated \cite{Dijkstra}; the extra entropy of a crystal of orientationally
disordered dimers, Fig. \ref{conf}(c), could easily overcome the free-energy balance with a
crystal of tetramers in some density range). The initially crystalline configuration
transforms into a fluid tetratic phase at $\eta\simeq 0.86$ (see Fig. \ref{ops1}). 
An interesting point is that this
tetratic phase is slightly different from the one obtained by compression of the isotropic phase,
Fig. \ref{ops}.
Clearly a relatively small system of hard right triangles is very prone to developing
metastable phases with a structure sensitively dependent on the initial configuration
and the path followed by the simulation. 

%

\begin{table}
\begin{tabular}{|c|c|c|c|}
\hline\hline
	{\bf phase} & {\bf uniaxial} & {\bf tetratic} & {\bf octatic}\\
	\hline
	$\eta$ & $0.820\pm 0.004$ & $0.822\pm 0.004$ & $0.790\pm 0.003$\\
\hline\hline
\end{tabular}
	\caption{\label{table2} Average densities of phases at pressure $\beta pa=13.5$
	according to $NpT$ simulations. Each $10^6$ MC step-simulation starts from a 
	different configuration
	(uniaxial, tetratic or octatic) obtained in different $NVT$ runs.}
\end{table}

As a final analysis, we performed $NpT$ simulations at a fixed value of pressure, $\beta pa=
13.5$ (which should be right at the liquid-crystal density interval \cite{Dijkstra}), starting from configurations
with uniaxial, tetratic, and octatic symmetries. The average densities obtained are given in Table \ref{table2}.
Note that $\eta_{\rm octatic}<\eta_{\rm uniaxial}<\eta_{\rm tetratic}$. As a general rule
one expects the stable phase to have the highest density at fixed pressure. Of course,
from a thermodynamic point of view, this is no proof that the T phase is more stable 
than the others, but it certainly gives a hint. Interestingly the O phase is considerably less
compact than the T phase, with the uniaxial N phase quite close to the latter.

%
%

From the above discussion, we cannot strictly 
discard the existence of a purely O phase between
the isotropic fluid and the crystal of a system of hard right-triangle particles. However, 
in the region where the O phase could be stable, a competing phase with 
clear tetratic symmetry also arises in simulations (starting from the crystal
of tetramers). The T phase is more
plausible than the O phase from the point of view of particle configurations.
Even though the $NpT$ simulations indicate that the T phase might be more stable, a definite answer can only
come from free-energy calculations. 

\begin{figure}
\epsfig{file=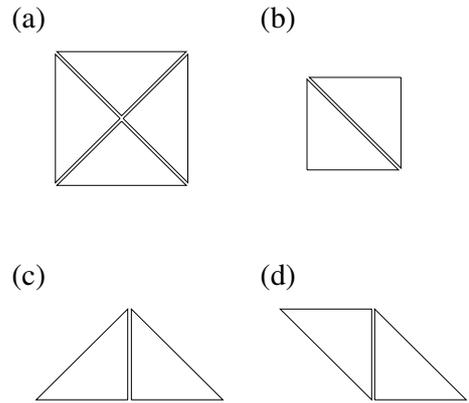,width=2.5in}
	\caption{\label{clust} Possible two- and four-particle configurations for hard
	right triangles. (a) Tetramer. (b) Square dimer. (c) Triangular dimer. (d) Rhomboidal
dimer.}
\end{figure}

The different symmetries
obtained depend very sensitively on the starting configuration and the path, either
compression or expansion, followed in the simulation runs. 
Obviously the high packing and the two dimensionality of the system
causes particles to have a low rotational diffusion and to find 
difficulties to explore different local symmetries.
An additional effect is the clustering tendencies of particles with right linear sectors.
A recent study relates the formation of
clusters of polyhedral particles with chemical bonding \cite{Glotzer2}. Local particle
configurations vary in type according to the global symmetry of a phase. For hard right
triangles several clusters can be identified, see Fig. \ref{clust}. In the
T phase we expect a high fraction of dimers (two triangles or `monomers' in close
contact along their long or short sides, forming groups with square and triangular shape, 
respectively) and tetramers [four close monomers in configurations as shown in 
Fig. \ref{clust}(a)], and
a low proportion of dimers forming a rhomboid. The latter are present in the isotropic
and octatic phases, which means that a compression of the isotropic phase will lead to an octatic 
texture or frustrated tetratic phase. 
In turn, a T phase may also form in the absence of tetramers, giving rise to 
different tetratic configurations with a history-dependent structure. 

In any case, our simulations seem to rule 
out the existence of the uniaxial N phase predicted by the standard SPT-based DFT
or the extended SPT theory including the third virial coefficient.
The intermediate phase between the isotropic and the crystal phases most 
likely exhibits strong tetratic and octatic correlations, none of which are detected in the DFT calculations.
However, a final answer will have to wait for a more detailed simulation study including
free-energy calculations, and the stability of a uniaxial N phase in a limited density
interval cannot be ruled out.

\section{Conclusions}
\label{conclusions}

In the present study we have shown that a SPT-based DFT incorporating the exact second and third virial coefficients  
is not capable of predicting stable T and O liquid-crystal phases in a fluid of hard right isosceles triangles. 
In addition, the orientation distribution function $h(\phi)$ of the only stable phase predicted, 
namely the uniaxial phase, has no signature of T or O correlations. However these symmetries were found in the MC 
simulations of Ref. \cite{Dijkstra} and in our own simulations, as shown by the orientational distribution function 
$h(\phi)$ and the order parameters $Q_k$ ($k=4,8$) in some range of packing fractions. 

Our results imply that the stable symmetries of the fluid of hard right triangles are driven by 
particle correlations involving more than three particles. In view of the practical impossibility to
implement higher-order correlations in a DFT model, the results of the present article
point to the necessity to formulate some new theoretical description, still within the framework of DFT,
but different from its standard form, to describe the phase behavior of right isosceles triangles. 
Indeed virtually all theoretical DFT models constructed so far for the description of uniform phases of two-dimensional 
or three-dimensional anisotropic particles are ultimately based on the knowledge of the second virial coefficient.
These models have been successfully applied to the description of liquid-crystal symmetries
for a vast variety of hard particles, and in general they correctly predict the symmetry of these 
phases and their stability ranges qualitatively, and even quantitatively in some cases.
The present fluid is, to our knowledge, the first example where the standard theoretical 
tool used by the liquid-crystal community since Onsager formulated his theory breaks down.

However, there are different avenues to remedy the deficiencies of the standard 
theory, still within the framework of DFT. 
We suggest that clustering (or self assembling) of particles may be a fruitful
view to overcome the difficulties posed by an approach based on a virial expansion. Particle shape
alone is enough to identify particularly stable local particle configurations, which can be taken as
entities upon which to formulate a DFT model. This approach would automatically incorporate the
important correlations through the very identification of the possible clusters, the distribution
of which would be obtained in a thermodynamically consistent way, providing a
mechanism to explain the peculiar (four- or eightfold) orientational symmetries
of the liquid-crystal phases of hard right triangles.
For example, the presence of a large amount of dimers and tetramers 
forming square-like super-particles and both arranged in a T-like 
configurations can lead to monomer axes, depending on their relative fractions, that orient parallel 
to four or eight, approximately equivalent, directors. Consequently the angular distribution functions 
of monomers could approach the symmetries $h(\phi)=h(\phi+n\pi/4)$ ($n=1,2$).

It is reasonable to appeal to the clustering effect to explain why the T and O phases, 
observed in simulations, cannot be stabilized by the usual  
implementation of DFT. A more realistic model to account for clustering is to treat the fluid as a polydisperse 
mixture of clusters of different sizes and shapes. 
An internal energy for clusters, in the line of standard 
models for associated fluids, should be introduced. 
A study of the relation between phase symmetry and particle clustering, and the formulation of
a cluster model, which we believe is a promising research line, is left for the future.

To end, we would like to comment on two distinct extensions of the problem.
First, our DFT study is restricted only to uniform phases as by construction these
are the liquid-crystal orientational symmetries that the present DFT can predict.
The study of non-uniform phases can be tackled with a theory optimized for
inhomogeneous density profiles. One of the promising versions of DFT
is that constructed from fundamental measure theory
(see for example Ref. \cite{Wittmann}). This theory was formulated for any hard particle geometry. 
Even though the theory is computationally expensive, it should be possible to 
fix different crystalline lattices with T symmetry, for example the ones
formed by tetramers or by randomly oriented dimers,
and minimize the functional with respect to the density profile. The relative stability 
between different crystalline phases or between the fluid and the crystal could then be studied
and compared with the predictions of MC simulations \cite{Dijkstra}.

Finally, one might wonder about the effect of polydispersity in the present system. A 
perfect nematic phase of monomers 
and also a perfect tetratic phase of tetramers under compression lead to perfect plane tiling
at close packing, with the latter 
having greater orientational entropy and consequently lower free energy. If polydispersity in the opening 
angle is present, it could affect the stability of these two phases. In particular the balance between the 
orientational and configurational entropy can severely be affected in the T phase of tetramers. 
The stability of the crystal phase of tetramers should be more affected
by polydispersity because polydisperse monomers cannot fit into 
identical tetrameral units to form a crystal. However the effect could be less for the uniaxial crystal formed 
by monomers, possibly resulting in a stable uniaxial N phase at lower densities.

\acknowledgements

Financial support under grant FIS2017-86007-C3-1-P
from Ministerio de Econom\'{\i}a, Industria y Competitividad (MINECO) of Spain.
Y. M.-R. acknowledges the support from Grant No. PGC2018-096606-B-I00 (MCIU/AEI/FEDER, UE).



\begin{thebibliography}{30}
	\bibitem{Chaikin} K. Zhao, C. Harrison, D. Huse, W. B. Russel, and P. M. Chaikin,
        Phys. Rev. E {\bf 76}, 040401(R) (2007).
\bibitem{Mason} K. Zhao, R. Bruinsma, and T. G. Mason, Nat. Commun. {\bf 3}, 801
(2012).
\bibitem{Mason2} K. Zhao, R. Bruinsma, and T. G. Mason, Proc. Natl. Acad. Sci.
        USA {\bf 108}, 2684 (2011).
\bibitem{Escobedo} C. Avenda\~no and F. A. Escobedo, Soft Matter {\bf 8}, 4675 (2012).
\bibitem{Schlacken}  H. Schlaken, H.-J. M\"ogel and P. Schiller, Mol. Phys. {\bf 93}, 777 (1998).
\bibitem{Frenkel}K. W. Wojciechowski and D. Frenkel, Comp. Meth. Sci. Tech.
        {\bf 10}, 235 (2004).
\bibitem{MR} Y. Mart\'{\i}nez-Rat\'on, E. Velasco, and L. Mederos, J. Chem. Phys. {\bf 122}, 064903 (2005).
\bibitem{Donev} A. Donev, J. Burton, F. H. Stillinger, and S. Torquato, Phys.
        Rev. B {\bf 73}, 054109 (2006).
\bibitem{Selinger} J. Gen and J. Selinger, Phys. Rev. E {\bf 80}, 011707 (2009).
\bibitem{Sabi} S. Mizani, P. Gurin, R. Aliabadi, H. Salehi and S. Varga, J. Chem. Phys. {\bf 153}, 034501 (2020).
\bibitem{Dijkstra} A. P. Gantapara, W. Qi, and M. Dijkstra, Soft Matter {\bf 11}, 8684 (2015).
\bibitem{MR2} Y. Mart\'{\i}nez-Rat\'on, A. D\'{\i}az-De Armas and E. Velasco, Phys. Rev. E {\bf 97}, 052703 (2018).
\bibitem{Roth} R. Roth, K. Mecke, and M. Oettel, J. Chem. Phys. {\bf 136}, 081101 (2012).
\bibitem{Strandburg} K. J. Strandburg Rev. Mod. Phys. {\bf 60}, 161 (1988); 
	Erratum Rev. Mod. Phys. {\bf 61}, 747 (1989).
\bibitem{Mak} C. H. Mak, Phys. Rev. E {\bf 73}, 065104(R) (2006).
\bibitem{Engel} M. Engel, J. A. Anderson, S. G. Glotzer, M. Isobe, E. P. Bernard, 
	and W. Krauth, Phys. Rev. E  {\bf 87}, 042134 (2013).
\bibitem{Iyetomi} H. Iyetomi and P. Vashishta, Phys. Rev. A {\bf 40}, 305 (1989).
\bibitem{Schiper} A. G. Schlijper and R. Kikuchi, J. Stat. Phys. {\bf 61}, 143 (1990).
\bibitem{Menon1} V. Narayan, N. Menon and S. Ramaswamy,  J. Stat. Mech. P01005 (2006).
\bibitem{Dani}  T. M\"uller, D. de las Heras, I. Rehberg and K. Huang, Phys. Rev. E {\bf 91}, 062207 (20015).
\bibitem{GP1} M. Gonz\'alez-Pinto,  F. Borondo, Y.  Mart\'{\i}nez-Rat\'on  E. Velasco, Soft Matter {\bf 13},  2571 (2017).
\bibitem{GP2} M. Gonz\'alez-Pinto, J. Renner, D. de las Heras, Y.  Mart\'{\i}nez-Rat\'on and
        E. Velasco, New J. Phys. {\bf 21} 033002 (2019).
\bibitem{Menon2} L. Walsh and N. Menon, J. Stat. Mech. 083302 (2016).
\bibitem{Ariel} A. D\'{\i}az-De Armas, M. Maza-Cuello, Y. Mart\'{\i}nez-Rat\'on and E. Velasco,
        Phys. Rev. Research {\bf 2}, 033436 (2020).
\bibitem{Arts} R. Wittmann, L. B. G. Cortes, H. L\"owen and D. G. A. L. Aarts, Nat. Commun. {\bf 12}, 623 (2021).
\bibitem{Mason3}Z. Hou, Y. Zong, Z. Sun, F. Ye, T. G. Mason, and K. Zhao, Nat. Commun. {\bf 11}, 2064 (2020).
\bibitem{MR3} Y. Mart\'{\i}nez-Rat\'on and E. Velasco,  Phys. Rev. E {\bf 102}, 052128 (2020).
\bibitem{Glotzer}J. A. Anderson, J. Antonaglia, J. A. Millan, M. Engel, and S. C.
        Glotzer, Phys. Rev. X {\bf 7}, 021001 (2017)
\bibitem{Onsager} L. Onsager, Ann. N.Y. Acad. Sci. {\bf 51}, 627 (1949).
\bibitem{Lee} S.-D. Lee, J. Chem. Phys. {\bf 87}, 4932 (1987).
\bibitem{MR1} Y. Mart\'{\i}nez-Rat\'on, E. Velasco, and L. Mederos, J. Chem. Phys. {\bf 125}, 014501 (2006).
\bibitem{Glotzer2} E. S. Harper, G. van Anders and S. C. Glotzer, PNAS {\bf 116}, 16703 (2019).
\bibitem{Talbot} G. Tarjus, P. Viot, S. M. Ricci, and J. Talbot, Mol. Phys. {\bf 73}, 773 (1991).
\bibitem{Rigby} M. Rigby, Mol. Phys. {\bf 78}, 21 (1993).
\bibitem{Wittmann} R. Wittmann, C. E. Sitta, F. Smallenburg, and H. L\"owen, J. Chem. Phys. {\bf 147} (2017).
\end{thebibliography}
\end{document}